\documentclass[sigconf]{acmart}
\AtBeginDocument{%
  }

\renewcommand\footnotetextcopyrightpermission[1]{}

\copyrightyear{2025}
\acmYear{2025}
\setcopyright{rightsretained}
\acmConference[TEI '25]{Nineteenth International Conference on Tangible, Embedded, and Embodied Interaction}{March 4--7, 2025}{Bordeaux / Talence, France}
\acmBooktitle{Nineteenth International Conference on Tangible, Embedded, and Embodied Interaction (TEI '25), March 4--7, 2025, Bordeaux / Talence, France}\acmDOI{10.1145/3689050.3705992}
\acmISBN{979-8-4007-1197-8/25/03}

\usepackage{subcaption}
\usepackage{graphicx}

\begin{document}

\title[Mixed or Misperceived Reality?]{Mixed or Misperceived Reality? Flusserian Media Freedom through \textit{Surreal Me}}

\author{Aven-Le ZHOU}
\email{aven.le.zhou@gmail.com}
\orcid{0000-0002-8726-6797}
\affiliation{%
  \institution{The Hong Kong University of Science and Technology (Guangzhou)}
  \city{Guangzhou}
  \state{Guangdong}
  \country{P.R.China}
}

\author{Lei XI}
\email{xiju524954135@gmail.com}
\orcid{0009-0000-1802-6901}
\affiliation{%
  \institution{China Academy of Art}
  \city{Hangzhou}
  \state{Zhejiang}
  \country{P.R.China}
}

\author{Kang Zhang}
\email{kzhangcma@hkust-gz.edu.cn}
\orcid{0000-0003-3802-7535 }
\affiliation{%
  \institution{The Hong Kong University of Science and Technology (Guangzhou)}
  \city{Guangzhou}
  \state{Guangdong}
  \country{P.R.China}
}

\begin{teaserfigure}
  \centering
  \includegraphics[width=\textwidth]{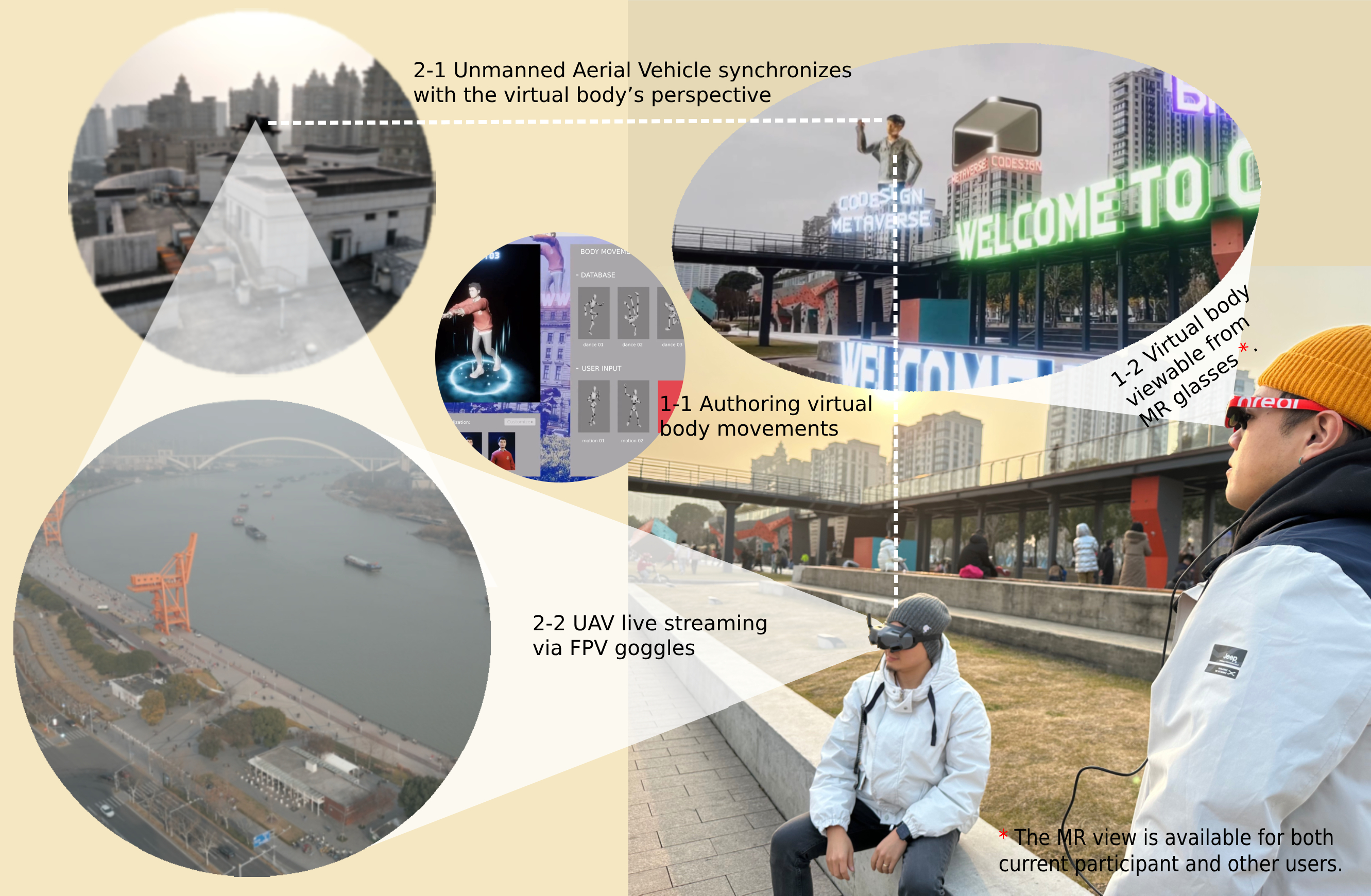}
  \caption{Two-phase virtual embodying experience of ``Surreal Me:'' The participant controls and authors the movement of their virtual body in MR, which is projected to the real world and viewable from MR devices. An Unmanned Aerial Vehicle synchronizes with the virtual body's perspective, in First-Person View, to the participants through goggles.}
  \Description{Interactive MR experience provided by Surreal Me: The participant's virtual body, through MR, is projected to the real world and viewable from MR devices. An Unmanned Aerial Vehicle (UAV) synchronizes with the virtual body's perspective and provides a First-Person View (FPV) with real-time video immersion to the participants through goggles.}
  \label{fig: teaser}
\end{teaserfigure}

\renewcommand{\shortauthors}{Zhou et al.}

\begin{abstract}

This paper delves into Vilém Flusser's critique of media as mediators that distort the human perception of reality and diminish freedom, particularly within the Mixed Reality context, i.e., Misperceived Reality. It introduces a critical inquiry through \textit{Surreal Me}, which engages participants to experience a two-phase virtual embodying process and reveal the ``Misperceived Reality.'' The process examines the obfuscating nature of media; as the Sense of Embodiment inevitably breaks down, users can discover the constructed nature of media-projected reality. When users reflect on reality's authentic and mediated experiences in MR, this work fosters a critical discourse on Flusserian media freedom addressing emerging immersive technologies.

\end{abstract}

\ccsdesc[500]{Applied computing~Media arts}
\ccsdesc[500]{Computing methodologies~Mixed / augmented reality}

\keywords{Flusserian Media Studies, Flusserian Media Freedom, Mixed Reality, Misperceived Reality, Embodiment, Embodied Interaction}

\maketitle

\section{Introduction}


According to Czech-Brazilian philosopher Vilém Flusser, media fundamentally shape our perception of reality by acting as intermediaries between humans and the world \cite[p.~9]{flusser2000}. The world is not ``immediately accessible'' to humans, so they need to ``make it comprehensible'' through media. Humans invent the media to help them make sense of the chaotic world \cite{ieven2003} and orient themselves in it \cite[p.~10]{flusser2000}. However, the media do not always offer faithful representations of the world. Instead, they encode information in particular ways. Since humans often misperceive the world projected by media to be the world itself, media become ``screens'' or ``coverings'' of the world \cite[p.~2]{flusser2013}, namely, obstacles between humans and the world. This misperception hinders humans from orienting themselves in the world through media and leads to a loss of freedom. This pivotal role of media, as outlined by Flusser, is becoming more pronounced with the rise of immersive technologies.

The tech industry defines mixed reality (MR) as a technology that ``brings together the real world and digital elements,'' indicating that reality in MR is often misrecognized with reality in the real world. But it is indeed a representation of the world projected through the digital apparatus. Thus, a problem arises: just as humans often misperceive medium to be the real world (i.e., Misperceived Reality), humans tend to regard the technical images of MR as the real world, thus are in danger of losing freedom and becoming a function of apparatus. Our work mainly investigates  Mixed Reality as an emerging media, leading the user to recognize the ``Misperceived Reality'' and obscuration of the real world in the media, thereby paving the way for new avenues to achieve Flusserian freedom. 

To pursue the Flusserian concept of media freedom in the context of Mixed Reality, we deliberately reveal the ``Misperceived Reality'' in MR. Through a speculative design project - ``\textit{Surreal Me}\footnote{This project was previously entitled as \textit{Surrealism Me} and exhibited in 2024 ACM Siggraph DAC Online Exhibition - The Future of Reality: Post-Truths, Digital Twins, and Doppelgängers, see \url{https://dac.siggraph.org/artwork/surrealism-me/}},'' we work on users' perception of the MR-projected world and alienate to the real, aiming to create them a contemplative moment that the ``world'' and ``self'' in MR are not real but surreal. 

\textit{Surreal Me} envisions and empowers an interactive MR experience where the user can own a virtual body. It uses a two-phase virtual embodying process first to accomplish the user's Sense of Embodiment (SoE) and then manipulates the user's visual and auditory sensations, disrupting SoE to demonstrate that reality in MR is not real. In doing so, we inspire a concern for the Misperceived Reality, media dominance, and risk of freedom in the contemporary digital context of emerging immersive technologies.

\section{Literature Review}

\subsection{Flusserian (Loss of) Freedom}

According to Flusserian thought, freedom means mastering the media. Thus, it argues that humans risk losing their freedom when they misperceive the media as the real world itself \cite{ieven2003}. In the Flusserian view of media history, such loss of freedom occurs frequently. With traditional images like paintings and drawings, humans often overlook their orientational purpose and mistakenly view them as accurate depictions of the world \cite[p.~10]{flusser2000}. To overcome the obscuration of the world by traditional images, humans invent the written texts \cite[p.~74]{flusser1997}. However, since written texts are ``one step further away from concrete experience than images,'' they often lead to ``a symptom of a bigger alienation'' \cite[p.~25]{flusser2002}. 

When the domination of written texts peaked in the nineteenth century, technical images emerged, exemplified by the invention of photography \cite{ieven2003}. However, the apparatuses that make the technical images are often black boxes to the ordinary user. In other words, users often do not understand how apparatuses encode the world. Apparatuses cease to be instruments that make the world accessible, and humans ``finally become a function of the images they create'' \cite[p.~10]{flusser2000}. 

\subsection{Towards Flusserian Freedom}

The call for freedom is significant for Flusserian media theory \cite{ieven2003}. Flusser suggests that humans can still be free by mastering and re-instructing the media that dominate us \citetext{\citealp[p.~73]{flusser2011}; \citealp[pp.~81-82]{flusser2000}}. Flusser states freedom means ``playing against the apparatus'' \cite[p.~80]{flusser2000}. More specifically, it involves imposing human intentions into the program of apparatus, exhausting its potential \cite{poltronieri2014communicology}. In this process, the user consciously creates unpredictable information to the apparatus, adding things that are not in the program \cite[pp.~81-82]{flusser2000}. In doing so, the user is not merely a function of the apparatus or conditioned by the apparatus but vice versa, which opens the way to freedom. 

In the Flusserian view, ``playing against apparatus'' is a strategy towards freedom through disrupting the program of the camera aimed at making technical images represent the real world \cite{Lenot_2017}. An example of this strategy is the experimental photographer who captures unpredictable images by reversing the steps of the shooting process instead of recording the real world \cite{Lenot_2017} \cite[pp.~81-82]{flusser2000}. Other than imposing unpredictable human intentions, Flusser proposes ``exhausting the program of apparatus'' as another essential way to play against it; thus, humans can be free \cite [pp.~80-82] {flusser2000}. For instance, game playing is the typical practice of humans playing against apparatus \cite{poltronieri2014communicology}, in which the player tends to exhaust the program within the game by constantly undertaking new actions \cite[pp.~26-27]{flusser2000}. 

\subsection{Misperceived Reality in Mixed Reality}

Over the decades, technical images have evolved from photography to immersive technologies, including Virtual Reality (VR) and Mixed Reality (MR). However, writings \cite[pp.~39,148]{popiel2012vilem} that address Flusserian freedom and immersive technologies tend only to involve VR while not mentioning MR. Although defined differently in different contexts, the common underlying trait of MR is the combination of the real world with virtual objects \cite{Rokhsaritalemi_2020}. Currently, video-see-through technology is a primary means of building MR spaces by displaying real-world and virtual objects on Liquid-crystal Display (LCDs) \cite{Potemin_2018}. In this context, the ``physical reality'' that Mixed Reality claims to contain is merely a digital representation. Thus, a problem arises: is the understanding of the world and the reality in MR misperceived? 

\subsection{Sense of Embodiment}

For immersive technologies aimed at simulating the physical world \cite{Pavithra_2020}, the Sense of Presence (SoP) is one of the main elements that generate the impression of realness \cite{Pietraß_2018}. SoP implies the feeling of ``being there'' in a virtual environment, which is an ``individual impression `when multi-modal simulations (images, sound, haptic feedback, etc.) are processed by the brain and understood as a coherent environment in which we can perform some activities and interact' \cite{Gutiérrez_2023}'' and a ``tendency of participants to respond to virtual events and situations as if they were real'' \cite{Slater_2006}. 

Various research concludes that SOP is intrinsically linked to \cite{Jung_2018} and depending on \cite{forster2022we} the Sense of Embodiment(SOE). Kilteni and Slater's framework defines SOE as three aspects: (1) Agency, the sense of being the author of a body's movements; (2) Body Ownership, the feeling that the body is the source of experienced sensations; and (3) Self-Location, the sense of being located inside the body \cite{6797786}. In other words, the establishment of SOE leads to the Sense of Presence and creates realness of reality as ``being there.'' On the other hand, breaking the SoE means undermining the SoP and disrupting the impression of realness, i.e., Misperceived Reality. 

\subsection{Related Work}
 
Various previous works investigate Sense of Embodiment (SOE) in VR and AR \cite{genay2021virtual, 9495125, 7504682,10.1145/3025453.3025602,10.1145/3488560.3502194}. In research related to SOE and MR, some works primarily investigate gaming technology and others focus on telepresence or the embodied experiences across realities \cite{9757562, 9319120, 10.1145/3487983.3488304, 8466636}. All these works emphasize the simulation of the world in MR, which contains the misperception that confuses the user about the relationship between the media and the world. Through Surreal Me, we investigate how humans can realize the misperceived reality in MR, better understand the media, and access Flusserian media freedom in the context of MR.


\begin{figure}[h]
  \centering
  \begin{subfigure}{0.44\textwidth}
      \includegraphics[width=\textwidth]{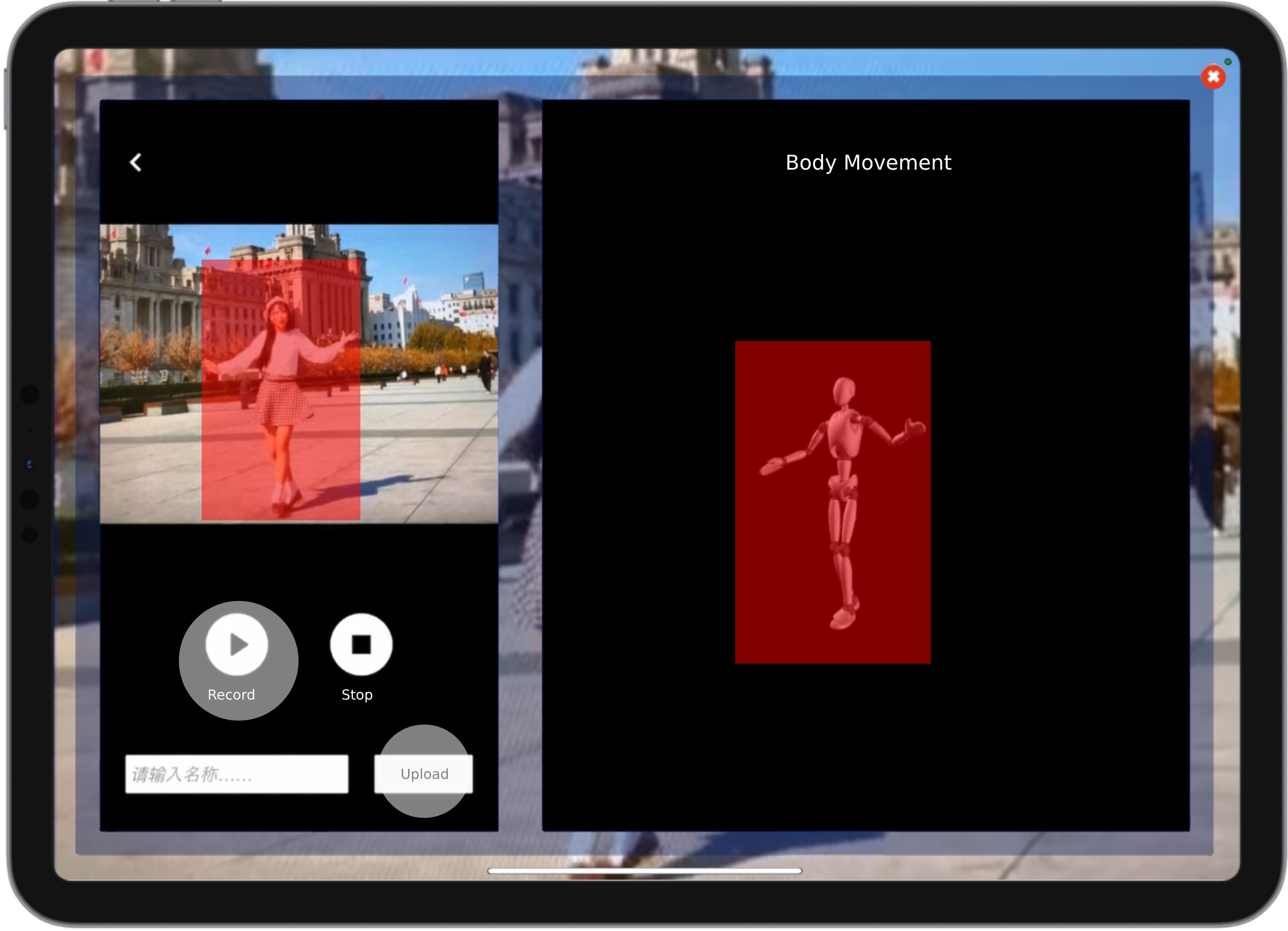}
      \caption{Capture body movement.}
      \label{fig:motion-left}
  \end{subfigure}
  \begin{subfigure}{0.44\textwidth}
      \includegraphics[width=\textwidth]{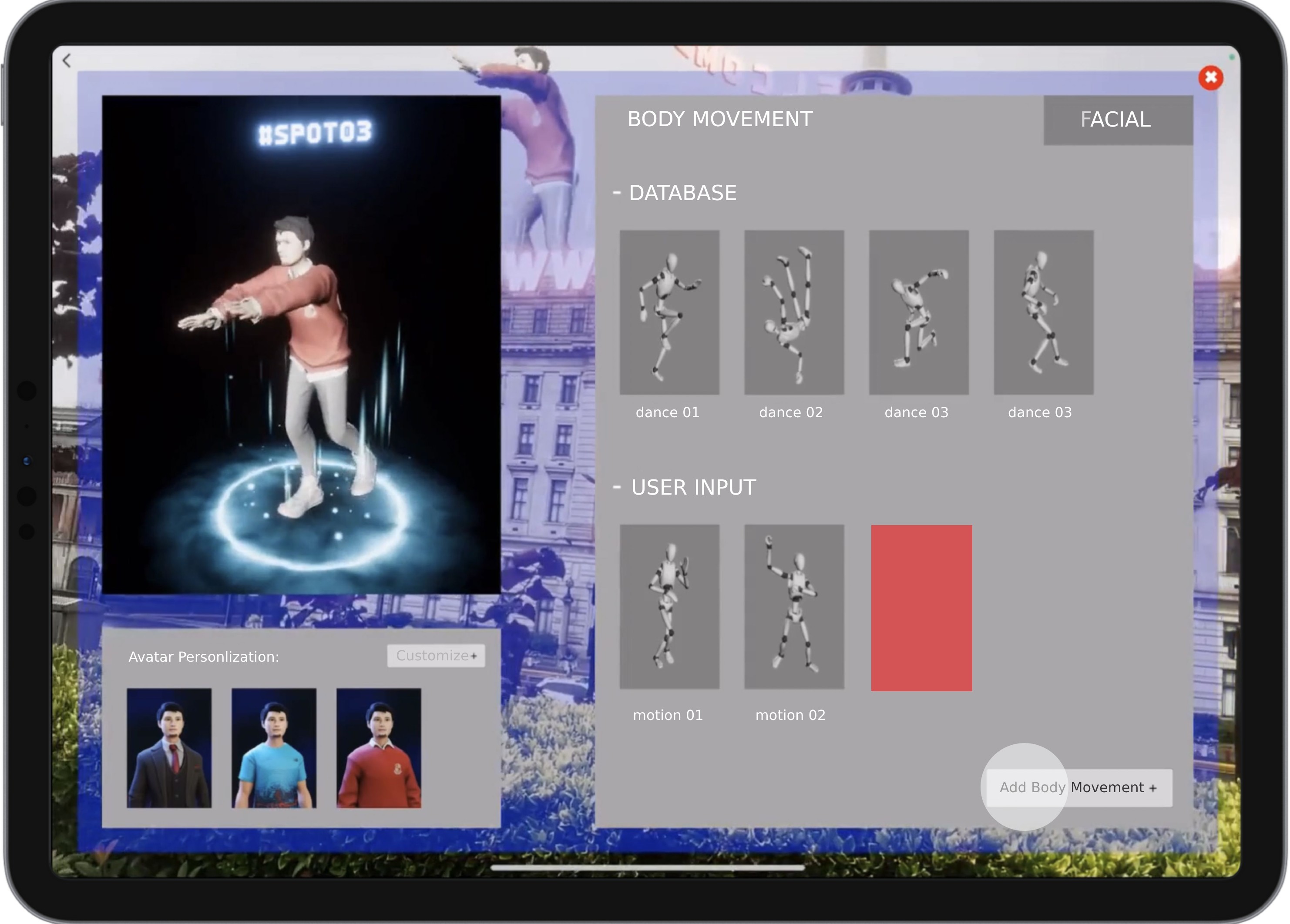}
      \caption{Choose motion from database.}
      \label{fig:motion-right}
  \end{subfigure}
  \begin{subfigure}{0.44\textwidth}
      \includegraphics[width=\textwidth]{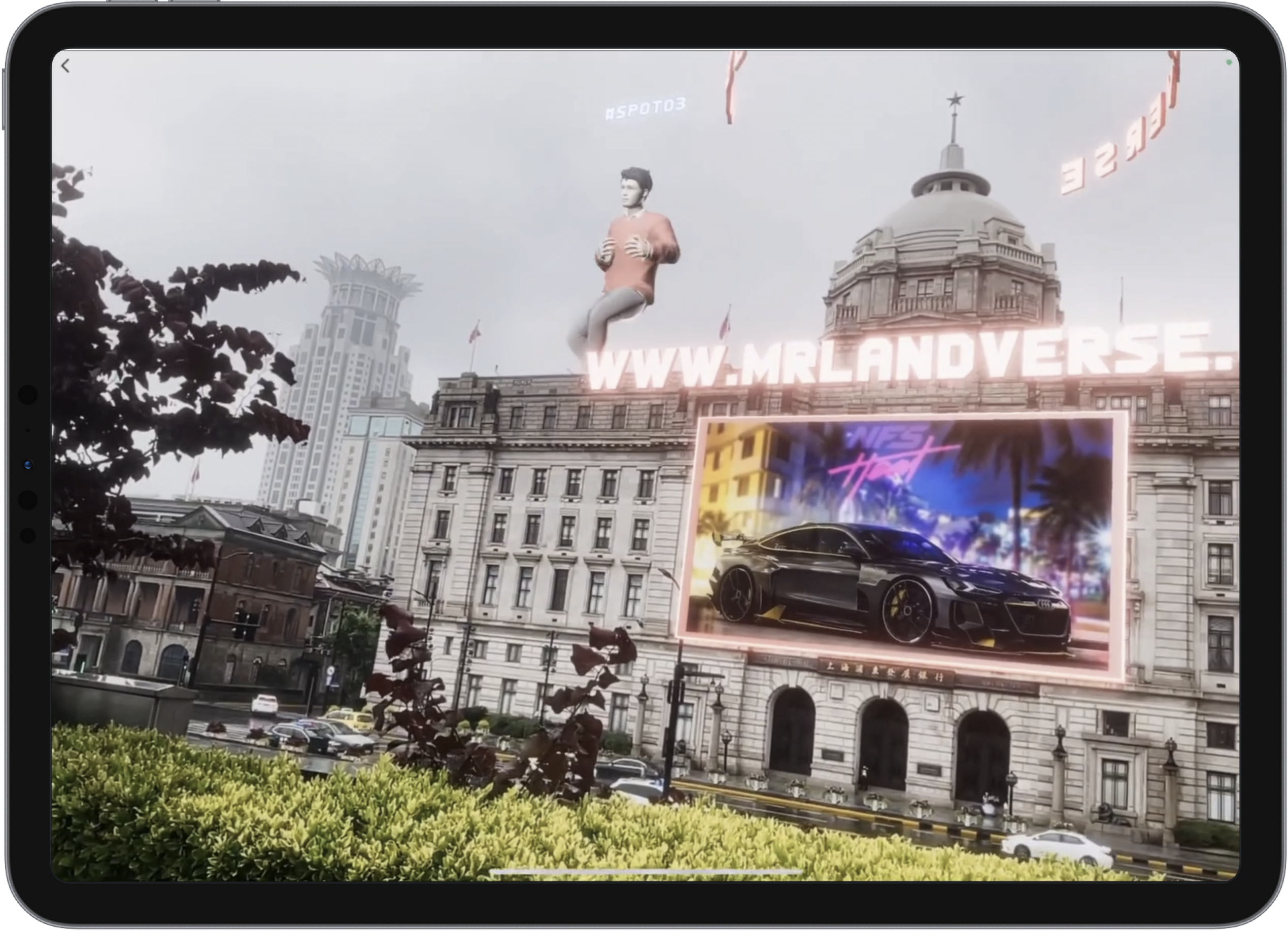}
      \caption{View virtual body in MR.}
      \label{fig:motion-ready}
  \end{subfigure}
  \caption{Through \textit{Surreal Me}'s customized software, users can manipulate their virtual body's motion and view it in MR, thus achieving the feeling of ``being the author of virtual body's motion.''}
  \label{fig:motion}
  \end{figure}
  
\section{Surreal Me}

\begin{figure*}[t!]
  \includegraphics[width=\textwidth]{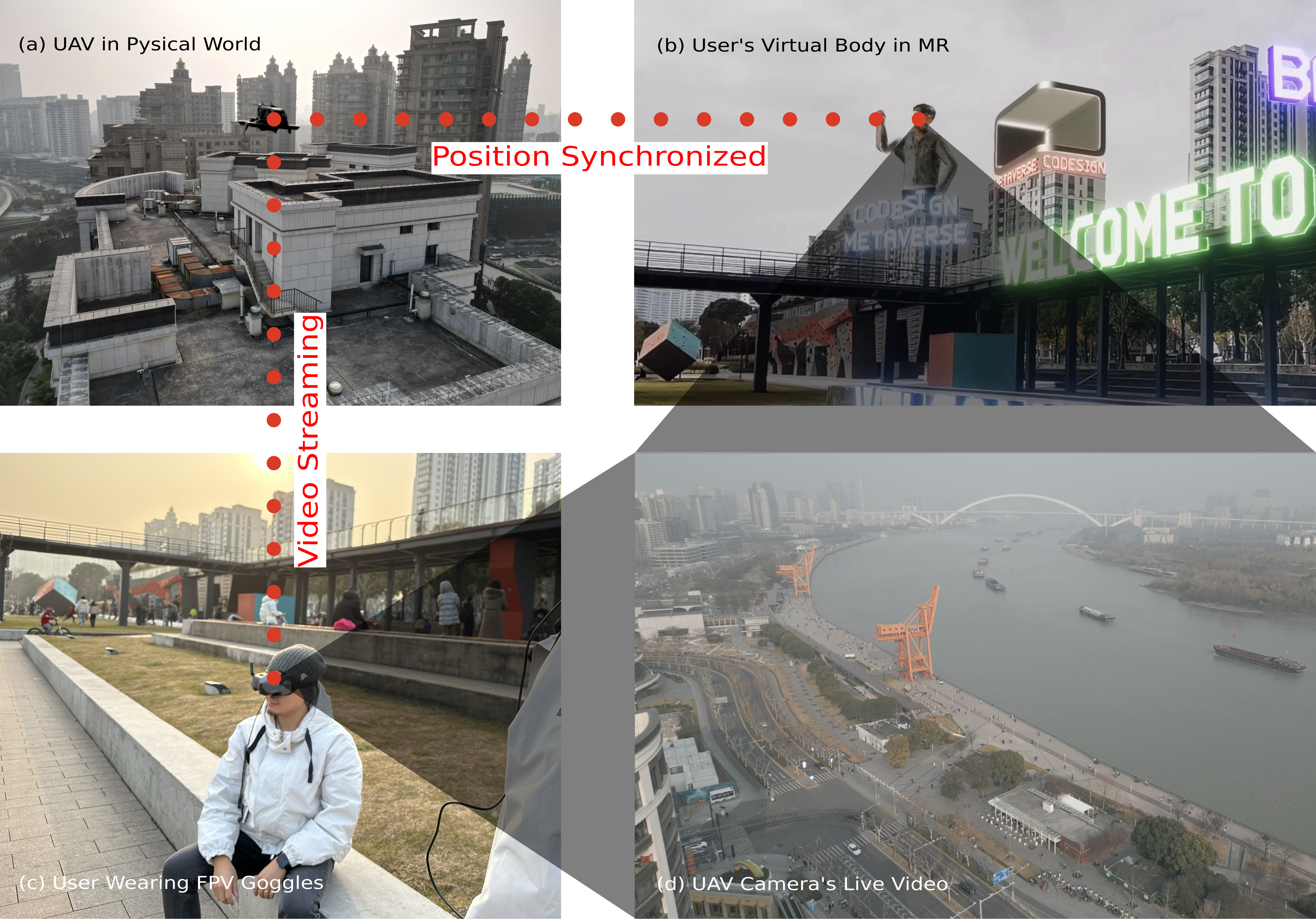}
  \caption{Through a First-Person View goggles and connected UAV, users can experience sensations their virtual body perceives in real-time and feel being located inside the body.}
  \label{fig:vision_modified}
\end{figure*}

To critically explore the ``Misperceived Reality'' in Mixed Reality, we present \textit{Surreal Me}, a speculative design project primarily investigating an interactive virtual embodying experience in MR following the Sense of Embodiment (SOE) framework. Through \textit{Surreal Me}, users first construct their SOE in MR, i.e., feel like they own a virtual body. Then, \textit{Surreal Me} disrupts the SOE and lets users know the virtual body is ``surreal me,'' a not-real experience, to create a contemplative moment for users. Following the lead of Kilteni and Slater's SOE framework, we design this process as two phases: (1) authoring the virtual body's movement and (2) feeling the virtual body's sensations and being inside. 

\subsection{Authoring Virtual Body's Movement}  
Fig. \ref{fig:motion-left} shows how users can move their bodies in front of the camera and capture their physical movement. In Fig.\ref{fig:motion-right}, they select either the motion they created or another one from \textit{Surreal Me}'s database as their preferred motion of the virtual body. The database consists of AI-generated motions and others created by users who granted permission to \textit{Surreal Me}. This enables users to ``be the author of the movement of their virtual body'' and is the first layer of the SOE experience. The virtual body is projected into the world and viewable from MR devices. Fig.\ref{fig:motion-ready} shows the \textit{Surreal Me} software interface where users can view the virtual body in MR. 



The \textit{Surreal Me} software shown in Fig.\ref{fig:motion} works almost identically on iOS devices and (android-based) MR glasses. Even though technically feasible, using MR glasses to capture the user's motion is impractical. In other words, it is only selecting motion and viewing on MR glasses, e.g., the man in the yellow hat in Fig.\ref{fig: teaser} was viewing the virtual body on top of the building. When combining two devices, \textit{Surreal Me} allows users to author the virtual body's motion in real-time, i.e., the user wears the MR glasses to view while capturing the motion with an iOS device running \textit{Surreal Me} software. The interface in Fig:\ref{fig:motion-left} continuously captures the user's physical motion and synchronizes it to their virtual body; the user can see and feel their virtual body following their physical body movements in real-time.

\subsection{Virtual Body's Sensations and Being Inside}
\textit{Surreal Me} employs an Unmanned Aerial Vehicle (UAV) to serve as the visual and auditory sensor of the user's virtual body. The UAV first flies to the virtual body's coordinates in the world (e.g., atop the building) and consistently synchronizes its position and perspective with the moving virtual body. The UAV in Fig.\ref{fig:vision_modified}(a) and the user's virtual body in Fig.\ref{fig:vision_modified}(b) share identical coordinates in the physical and virtual worlds. The UAV then streams the real-time video to the user through the FPV goggles. The user wearing goggles in Fig.\ref{fig:vision_modified}(c) can see the UAV camera's live video as in Fig.\ref{fig:vision_modified}(d). The user experiences the visual and auditory sensations of the virtual body, allowing them to immerse within their virtual body and feel they are located inside. 

Through the two phases of virtual embodying, \textit{Surreal Me} provides the user with a holistic experience of having the virtual body in MR following the definition of ``Sense of Embodiment.'' However, we intend to deconstruct the SOE to reveal the misperceived reality of MR. We design the ``spots'' (to place the virtual body, see Sec.\ref{sec:mr_solution}) mostly atop the buildings so that the UAV is often positioned at a high altitude when mimicking the perspective of the giant virtual body. Such a view is not usually accessible to users in everyday life, which creates an alienated feeling even though the view is realistic. The transition from the immersive experience provided by the UAV's perspective to the recognition of a simulated environment sparks a contemplative moment, leading to the realization that the MR world is a misperceived reality rather than real. In other words, users can realize their virtual body is not real but a ``surreal me.'' 

\begin{figure}[h]
  \centering
  \begin{subfigure}{0.45\textwidth}
      \includegraphics[width=\textwidth]{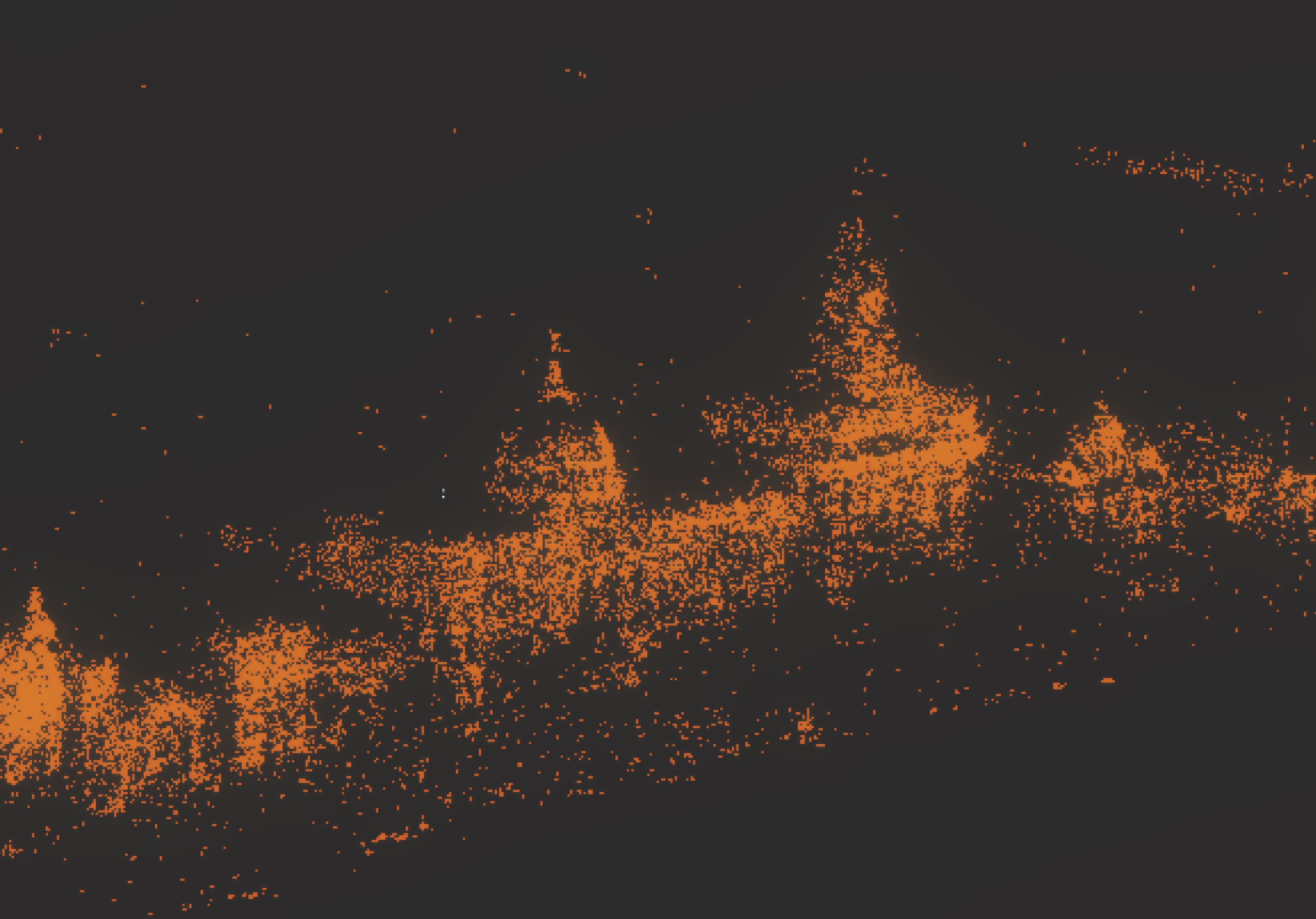}
      \caption{3D point clouds of selected Area of Interest (AOI).}
      \label{fig: MRtech-a}
  \end{subfigure}
  \hfill\hfill
  \begin{subfigure}{0.45\textwidth}
      \includegraphics[width=\textwidth]{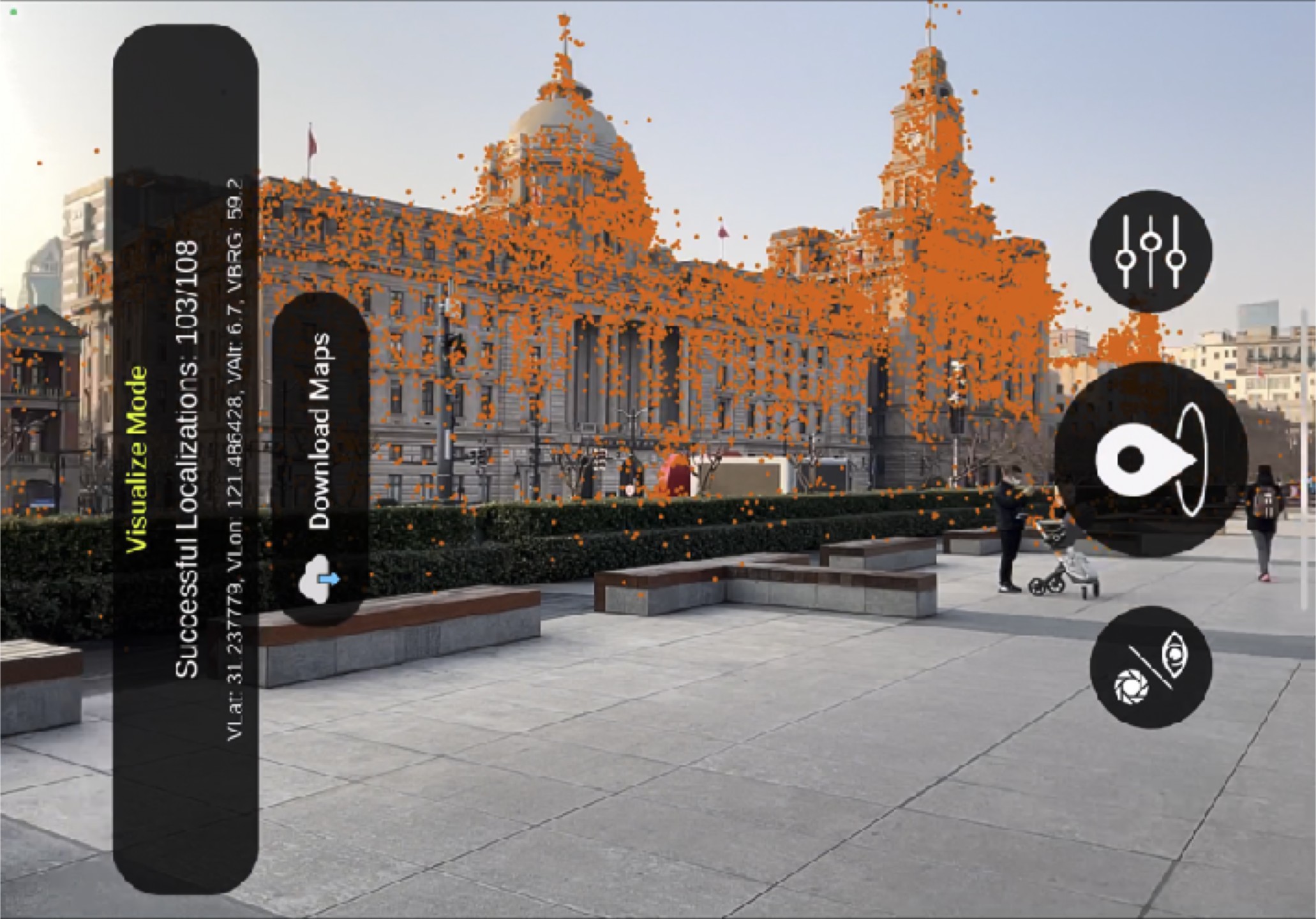}
      \caption{AOI identification with Visual Positioning System (VPS).}
      \label{fig: MRtech-b}
  \end{subfigure}
  \hfill\hfill
  \begin{subfigure}{0.45\textwidth}
      \includegraphics[width=\textwidth]{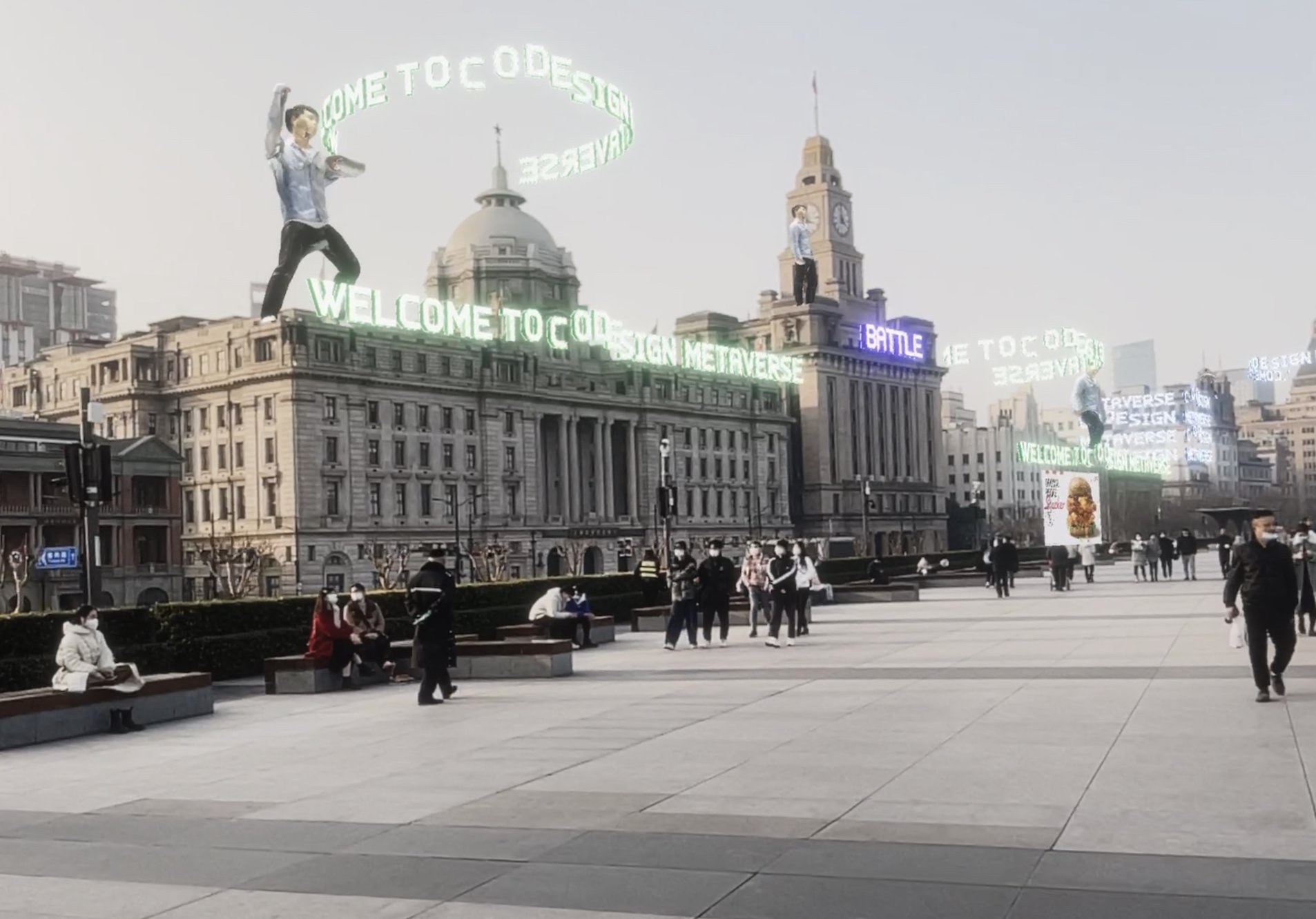}
      \caption{Virtual content anchored in the physical world as MR experience.}
      \label{fig: MRtech-c}
  \end{subfigure}
  \caption{MR solution applied in \textit{Surreal Me}.}
  \label{fig: MRtech}
  \end{figure}
  
\section{Experiment}
Setting up an MR experience mostly involves registering the physical site and placing digital content (in many cases, non-interactive ones), after which end users access and view the overlaid imagery through MR devices. In this artist-led inquiry, we have prepared and pre-registered several locations as the MR environment to introduce the users to virtual embodying quickly. 

\begin{figure*}[h!]
\centering
\begin{subfigure}{\textwidth}
\centering
  \includegraphics[width=0.8\textwidth]{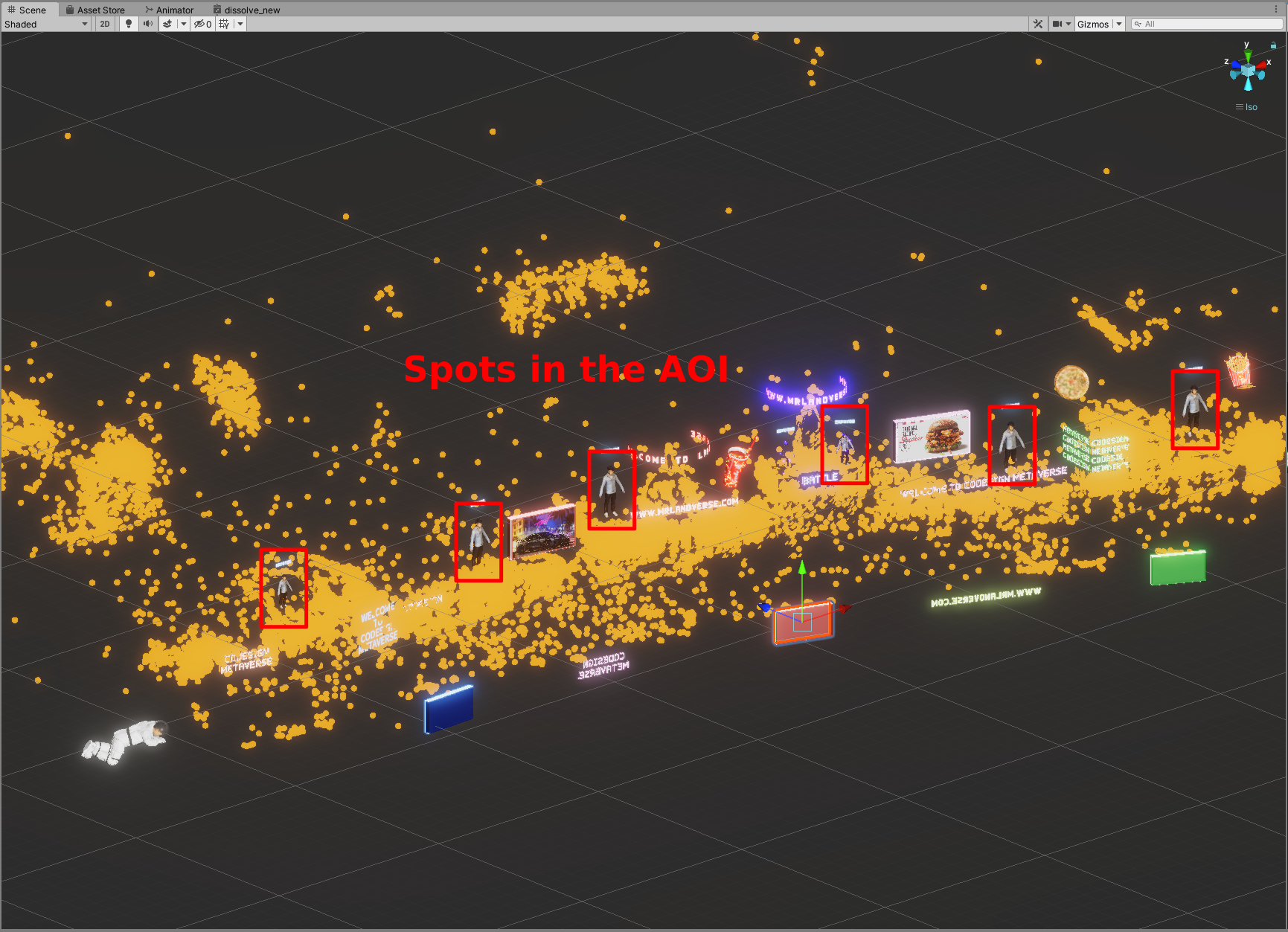}
  \caption{Point clouds with identified ``spots'' (screenshot from Unity).}
  \label{fig:place-left}
\end{subfigure}
\hspace{80em}%
\hspace{80em}%
\begin{subfigure}{\textwidth}
\centering
  \includegraphics[width=0.8\textwidth]{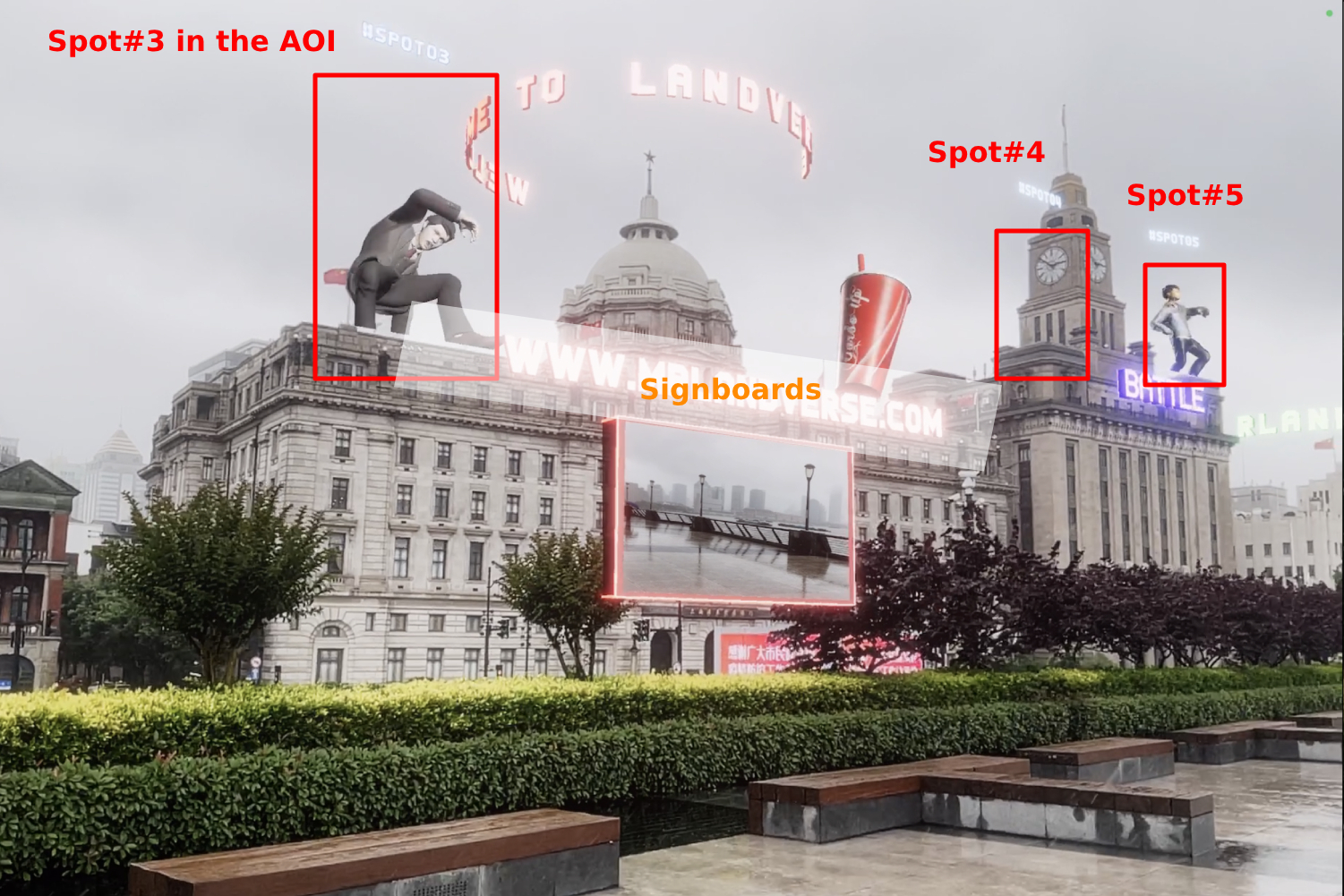}
  \caption{User's virtual body inside a ``spot'' (screenshot of \textit{Surreal Me}).}
  \label{fig:place-right}
\end{subfigure}
\caption{``Spots'' to place the virtual body.}
\label{fig:placement}
\end{figure*}

\subsection{MR Solution, Pre-registered AOIs and ``Spots''} \label{sec:mr_solution}
\textit{Surreal Me} employs Immersal SDK as the MR solution and integrates the Immersal Cloud Service to construct 3D point clouds of selected Areas of Interest (AOI) from site photos. These point clouds (as in Fig.\ref{fig: MRtech-a}) work with the Visual Positioning System (VPS) to provide high-accuracy localization and identification of the AOI (in Fig.\ref{fig: MRtech-b}). The combination of local SLAM and VPS creates the illusion that virtual content is anchored to the real world (as in Fig.\ref{fig: MRtech-c}). We have pre-registered several sites for the interactive MR experience and identified ``spots'' in the AOI that are ideal for placing and anchoring virtual bodies. Fig.\ref{fig:place-left} showcases how we have set up the ``spots'' in Unity about their positions. 

In \textit{Surreal Me}'s interface, the user can see the glowing label of ``spots'' in the MR environment and ``click'' on the spot to select it to place their virtual body. The click directs the user to Fig.\ref{fig:motion-right}, where the user can navigate through the database and apply a chosen motion to their virtual body and update it to the selected ``spot.'' Meanwhile, the ``add body movement'' button on the bottom right corner in Fig.\ref{fig:motion-right} leads to the capture motion interface in Fig.\ref{fig:motion-left}. Fig.\ref{fig:place-right} is a screenshot from \textit{Surreal Me} on an iPad, showing the perceptual illusion of the (virtual) body's presence within the MR environment. The demonstration scene includes additional decorative elements intended for a visually captivating presentation in MR; when placed on the cornice of the building facade, those glowing signboards can effectively conceal the misplacement of virtual content caused by the glitch of VPS and SLAM.

\subsection{Virtual Body}

\begin{figure*}[h]
  \centering
  \begin{subfigure}{0.525\textwidth}

      \includegraphics[width=\textwidth]{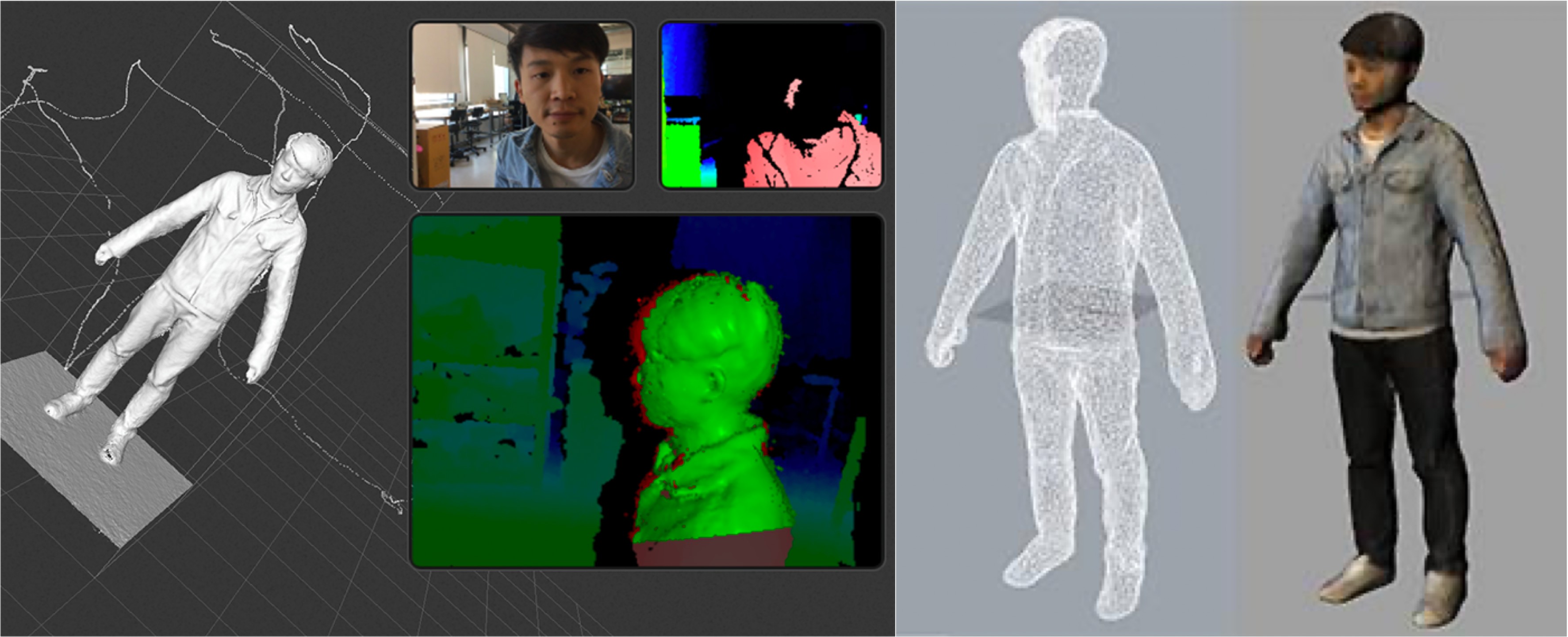}
       \caption{Workflow to scan the user with a Structure Sensor Pro, an iPad, and Skanect.}
      \label{fig: scan}
  \end{subfigure}
  \begin{subfigure}{0.465\textwidth}
      \includegraphics[width=\textwidth]{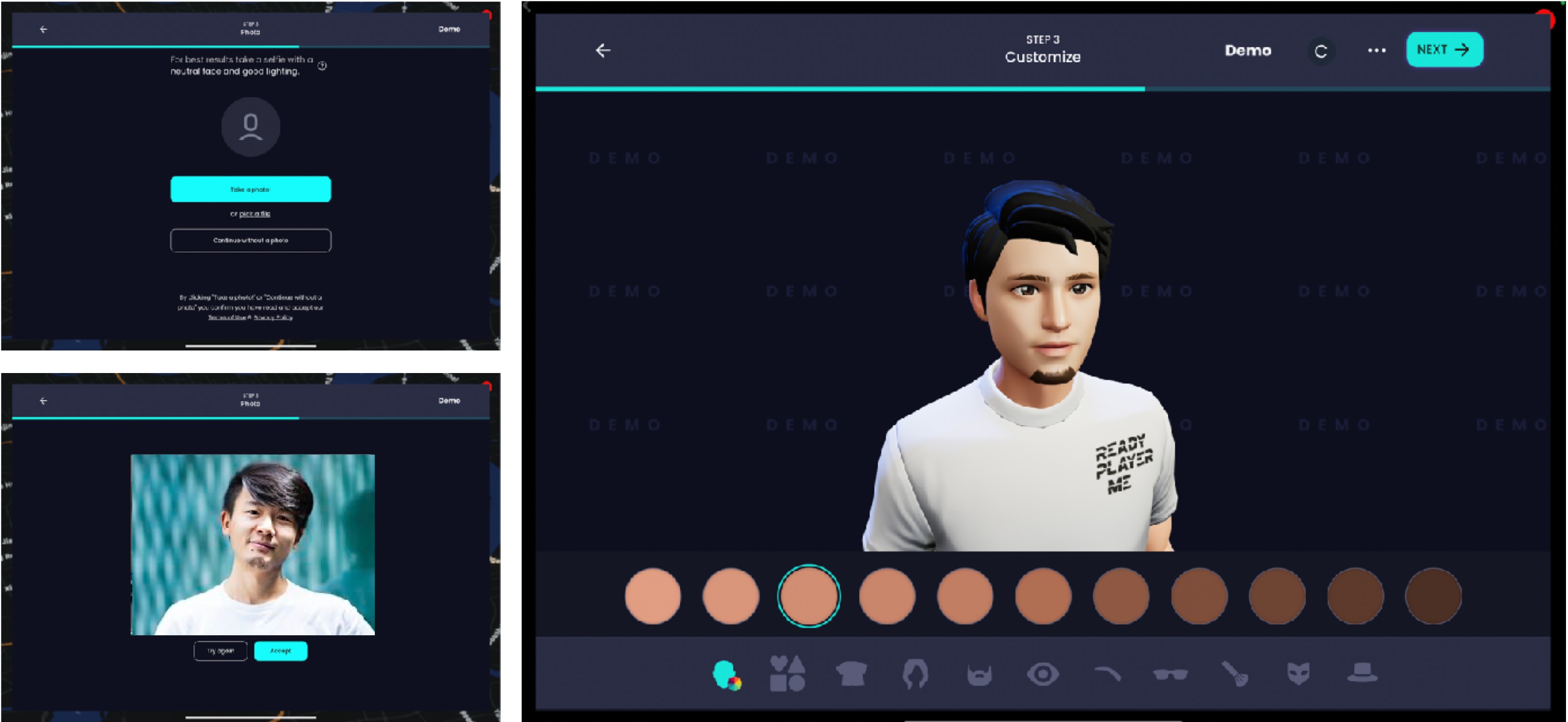}
      \caption{Integrated ``Ready Player Me'' interface to create an avatar for the virtual body.}
      \label{fig: palyer}
  \end{subfigure}
  \caption{Integrated ``Ready Player Me'' interface to customize an avatar as the virtual body.}
  \label{fig:avatarfig}
  \end{figure*}

In \textit{Surreal Me}, we have implemented two methods for the user to obtain their virtual body quickly. Fig.\ref{fig: scan} demonstrates the workflow to scan the user with a Structure Sensor Pro to create a realistic virtual body. Fig.\ref{fig: palyer} shows \textit{Surreal Me} integrates Ready Player Me Unity SDK to allow the user to customize a cartoon-style avatar as the virtual body. In our experiments, fewer participants chose the first method, but more preferred cartoon-style virtual bodies, which may indicate a favor to hide their identities from participants. 

\subsection{Virtual Embodying Experience and Interactions} \label{manipulation}
In \textit{Surreal Me}, we have employed Unity AR Foundation to capture the user's bodily movements, eventually supporting the user directly controlling their virtual body. Additionally, the experience can go beyond users posing and ``performing'' before the camera but allow them to upload video footage containing motion data to create movements. The user input motion, based on their permission, is archived in \textit{Surreal Me}'s database and available for other users to choose.


Moreover, through AI's generative capability, the motion database in \textit{Surreal Me} frees the virtual body's movement options from merely imitating users. We have initially incorporated an open-source body movement dataset into \textit{Surreal Me}'s database for users to choose from, such as different dancing. Additionally, \textit{Surreal Me} uses an open-source neural network \cite{9042236} to learn from a given training dataset (i.e., user-authored movements) and generate new motions and expand the database. As users constantly add their authored new motions, this ongoing experiment can evolve the AI algorithm by training it with the user's new inputs. Thus, the AI generative capability keeps growing along with the engagement of humans; so does the database and the motion options it may provide. In this case, the AI introduces a semi-autonomous agency and works collectively with the users to author their virtual body movements. 

\begin{figure*}[h]
\centering
\begin{subfigure}{0.425\textwidth}
  \includegraphics[width=\textwidth]{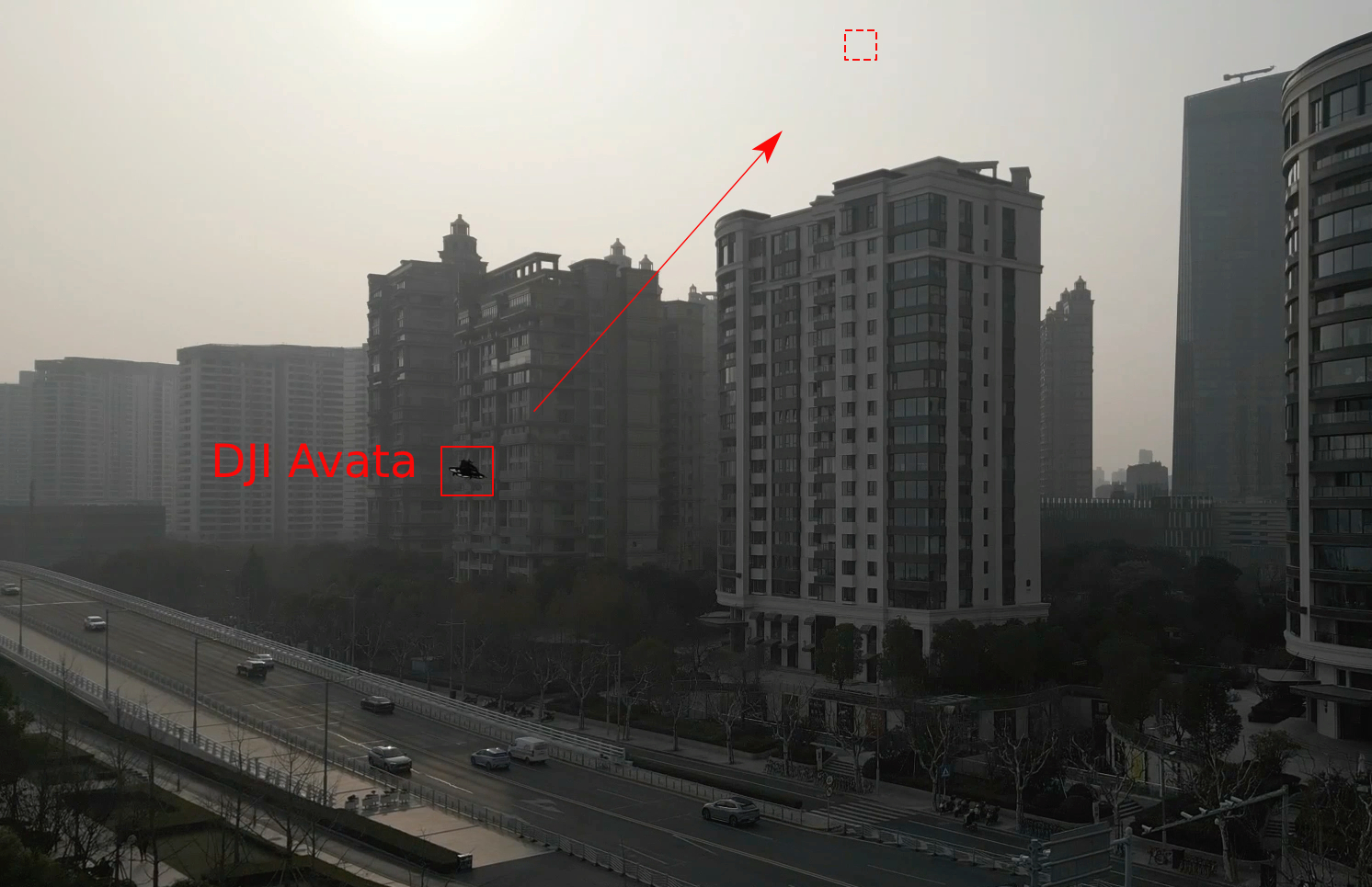}
  \caption{DJI Avata drone flying to virtual body.}
  \label{fig:droneflight}
  \end{subfigure}
\begin{subfigure}{0.55\textwidth}
  \includegraphics[width=\textwidth]{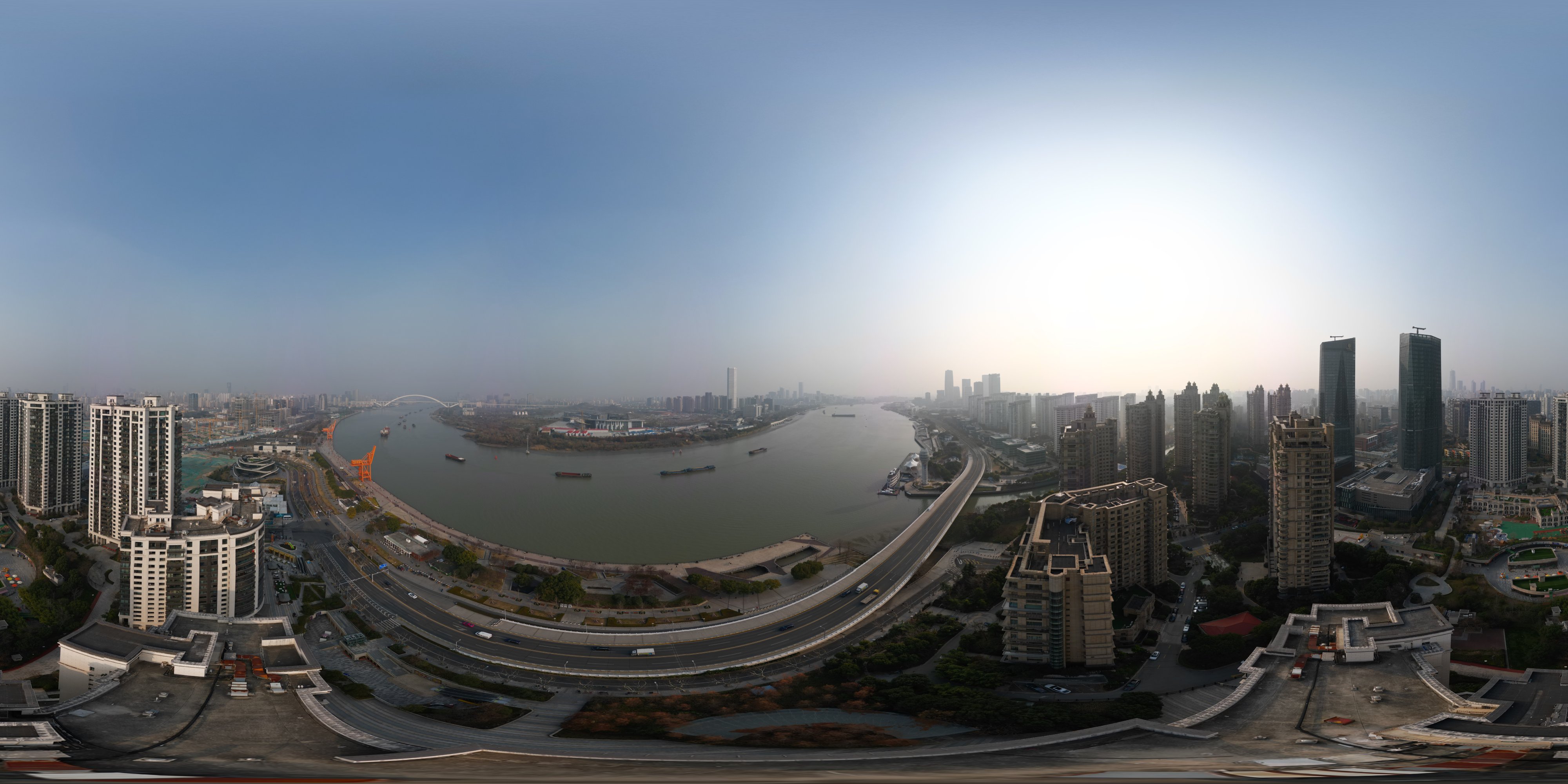}
  \caption{FPV view captured in real-time.}
  \label{fig:droneview}
  \end{subfigure}
\caption{Experiment with perceiving visual and auditory sensations of virtual body through drones and FPV goggles.}
\end{figure*}

The feeling of having a virtual body is present in the first phase of \textit{Surreal Me}, yet the feeling of owning it is insufficient. In the second phase, we have hacked a DJI Avata to execute the flying task (as in Fig.\ref{fig:droneflight}). It receives the coordinates of the ``spot'' that the virtual body occupies from \textit{Surreal Me}'s software and automatically flies to the destination. When the user wears the FPV goggles, the real-time view and sound (as in Fig.\ref{fig:droneview}) are live streamed from the UAV, enhancing their embodied experiences; the sense of ``being located inside the virtual body'' is present. Thus, the holistic Sense of Embodiment emerges — if not considering the unusual position of the virtual body and, correspondingly, the UAV's unfamiliar perspective.


\section{Summaries and Reflections}

\subsection{Misperceived Reality or Mixed Reality?}

\textit{Surreal Me} uses ``feeling the virtual body's sensations'' to create an ambivalence during the process: it offers enhanced virtual embodying experiences but causes an alienation from everyday life experiences. This ambivalence inspires the user to rethink the reality of MR. The experience eventually exposes the obfuscating effects of media by demonstrating how Sense of Embodiment breaks down, revealing the disparity between MR and reality. When the user contemplates the cause of this alienation, they can easily recognize the presence of the UAV as the mediator and realize that the seemingly real world in MR is not real but a world projected by the apparatus. As the SoE breaks down, \textit{Surreal Me} reveals the constructed fact of MR-projected reality, thus challenging human perception of media reality. It critically examines the nature of MR media and reflects on the media's dominant role, encouraging re-evaluation of the media's impact on human understanding of the world and advocating for Flusserian Freedom.

\subsection{Exhausting the Apparatus with AI}

Though it primarily investigates the ``Misperceived Reality'' in the MR context, this work re-situates the Flusserian idea of ``exhausting the program of apparatus'' in the pressing topic of the human-AI relationship. When the user chooses movements authored by others, the collective efforts of among users lightly touch the concept of ``exhausting the apparatus'' as in game playing. On the other hand, when the user-authored movements serve as a training dataset, AI learns from human input and generates more and more motions in the database for users to choose from. In this case, humans and AI collectively create body movements, ``exhausting the apparatus'' toward Flusserian freedom. 

\subsection{User Reflections}

In our preliminary pilot study, we engage participants with open-ended questions regarding their experience of embodiment,such as how they feel about owning the virtual body and their general feeling of experience across the two phases. More than half of those with MR experiences (four out of seven) could ``feel the virtual body'' in the first phase, and three of them mentioned ambivalence or disruption. One participant asked ``whether the purpose of this experiment is to construct or deconstruct the virtual embodying or both,'' and concluded it as ``to deconstruct it through constructing it.'' 


Participants with less exposure to MR technologies seem more into the illusive effects of MR and the FPV's immersive perspective; they are mainly into discussing the ``wow'' moments of seeing their virtual body as well as the excitement of the immersion within the FPV goggles. We tend to conclude that average users are still fascinated by the media and technology, especially if they haven't previously experienced them. The excitement of trying MR glasses and FPV goggles seems to conceal the potential contemplation moment. 

In the experiment, we noticed that almost the same group of participants who used the non-realistic avatar as their virtual body chose to use the existing motion rather than ``authoring'' in real time. Their choice of virtual body and movement could be relevant to their sense of identity, and we are curious whether it would differ across cultures. Furthermore, the idea of extending virtual embodiment experiences to individuals with disabilities or limited mobility surfaced multiple times, pointing to potential avenues for inclusive design enhancements. 



\section{Conclusion, Limitation, and Future Work}

\textit{Surreal Me} responds to Flusserian freedom discussed in media studies, especially the emerging Mixed Reality context. Most participants with prior experience achieved our desired ``breaking-down of SoE and MR'' and ``contemplative moment'' that reveals the ``Misperceived Reality.'' Meanwhile, we have provided our deductive reasoning for those who did not. It seems to be a match with the development of technology-mediated arts in general; when applying a new technical medium, the audience is always fascinated by the technology when it is new. They then turn to the concept of the work only when they are familiar with the technology. 

The pilot study's exploratory nature relied on anecdotal data, which, while providing initial insights, lacked the systematic rigor required for more definitive conclusions. The open-ended approach allowed for a rich tapestry of responses but did not facilitate easy quantification or cross-comparison of data. Additionally, the study's confinement to an Eastern context might not fully resonate with the Western-centric Flusserian theories, potentially biasing the applicability of the findings across different cultural backgrounds.

Building upon the foundational groundwork laid by "Surreal Me," future iterations will adopt a more structured approach to data collection. We intend to formulate a standardized set of interview questions that will enhance our ability to capture detailed and comparable data on user experiences of embodiment. This methodological enhancement will allow for a deeper and more quantifiable analysis of emerging themes.

Moreover, recognizing the diverse reactions based on cultural and experiential backgrounds, we plan to broaden the scope of our user testing. This expansion will include participants from varied cultures and those with disabilities, ensuring that our exploration into virtual embodiment is comprehensive and inclusive. Such an approach will validate and extend the Flusserian concepts discussed and broaden our findings' applicability to a global audience.

\bibliographystyle{ACM-Reference-Format}
\bibliography{sample-authordraft}

\appendix 

\end{document}